\documentclass[aps,superscriptaddress,floatfix,nofootinbib,twocolumn]{revtex4-1}
\usepackage{exscale}                  
\usepackage[intlimits]{amsmath}       
\usepackage{amsfonts}
\usepackage{amssymb,amscd}
\usepackage[dvips]{epsfig}                   
\usepackage{graphicx}
\usepackage{array}
\usepackage{wrapfig}
\usepackage{color}
\newcommand{\is}{\sum\!\!\!\!\!\!\!\int}

\newcommand{\beqa}{\begin{eqnarray}}
\newcommand{\eeqa}{\end{eqnarray}}
\newcommand{\beq}{\begin{equation}}
\newcommand{\eeq}{\end{equation}}

\newcommand{\Tr}{\mathrm{Tr}}

\definecolor{myred}{rgb}{1,0.8,0.8}

\begin{document}
\title{Propagators and phase structure of $N_f=2$ and $N_f=2+1$ QCD}
\author{Christian~S.~Fischer}
\affiliation{Institut f\"ur Theoretische Physik,
  Justus-Liebig-Universit\"at Gie\ss{}en,
  Heinrich-Buff-Ring 16,
  D-35392 Gie\ss{}en, Germany}
\affiliation{GSI Helmholtzzentrum f\"ur Schwerionenforschung GmbH, 
  Planckstr. 1  D-64291 Darmstadt, Germany.}
\author{Jan Luecker}
\affiliation{Institut f\"ur Theoretische Physik,
  Justus-Liebig-Universit\"at Gie\ss{}en,
  Heinrich-Buff-Ring 16,
  D-35392 Gie\ss{}en, Germany}
\date{\today}
\begin{abstract}

We investigate the phase structure of QCD at finite temperature and chemical potential 
by solving a coupled set of truncated Dyson-Schwinger equations for the quark 
and gluon propagator. In contrast to previous calculations we take into account
the full back-reaction of the quarks onto the Yang-Mills sector and we include
the effects of strange quarks. We discuss the resulting thermal mass of the 
unquenched gluon propagator and extract order parameters for the chiral and 
deconfinement transition from the quarks. Our result for the temperature 
dependence of the quark condensate at zero chemical potential agrees well with 
corresponding lattice calculations. We determine the phase diagram at finite chemical 
potential and find a potential critical endpoint at
$(\mu_q^{EP},T^{EP}) \approx (190,100)$ MeV.

\end{abstract} 

\maketitle

\section{Introduction}

The existence of a high temperature and/or density phase with quarks and gluons 
as thermodynamically active degrees of freedom has been a major prediction since 
the early days of QCD. Over the years this topic has received 
a lot of attention from both, the theoretical and experimental sides. Today, high
quality results are available from lattice simulations at zero chemical 
potential, see e.g. \cite{Borsanyi:2010bp,Bazavov:2011nk} and references therein. 
As it turned out, the low and high temperatures phases of QCD are not separated
by a phase transition but rather continuously connected. By now, lattice QCD has 
firmly established the notion of a crossover at physical quark masses and zero 
chemical potential. 

Unfortunately the situation is much less clear at (real) finite 
chemical potential, where lattice calculations are hampered by the notorious 
fermion sign problem. Although various extrapolation methods agree with each other 
at rather small chemical potential 
\cite{latticemu,deForcrand:2010he,Kaczmarek:2011zz,Endrodi:2011gv,Cea:2012ev}, 
regions in the $(T,\mu)$-plane with $\mu_q/T > 1$ are hardly accessible. Therefore, 
despite these efforts the basic structure of the phase diagram of QCD is not 
at all settled yet, see e.g. \cite{Weise:2012yv,Fukushima:2011jc} and references therein.

An alternative approach to the QCD phase diagram is functional methods, i.e.
the functional renormalization group (FRG) and Dyson-Schwinger equations (DSEs).
Naturally, these approaches have to rely on approximations in most 
cases\footnote{There are, however, limits where exact results are possible, 
see e.g. Ref.~\cite{Fischer:2009tn}.}, which are less controlled than those 
in lattice Monte Carlo simulations.
These approximations, however, are controlled by constraints such as symmetries 
and conservations laws and also by comparison of the results with corresponding 
ones from lattice calculations at zero and imaginary chemical potential. If this 
procedure is followed carefully, one may hope to obtain meaningful results also 
for the finite chemical potential region with $\mu_q/T > 1$.

One of the potential advantages of the FRG and DSE-approach to QCD as compared with 
models like the Nambu--Jona-Lasinio model (NJL) \cite{NJL}, its Polyakov loop extended versions
\cite{Fukushima:2003fw,Megias:2004hj,Ratti:2005jh} and the Polyakov loop extended quark-meson 
model (PQM) \cite{Schaefer:2007pw,Skokov:2010wb,Herbst:2010rf} is the direct
accessibility of the Yang-Mills sector. In the models, gluons are not active degrees 
of freedom and their reaction on the medium can thus neither be studied nor directly
taken into account. This is possible within the functional approaches to QCD. With 
functional renormalization group equations the gluons of quenched QCD where studied 
in Ref.~\cite{Fister:2011uw}, and QCD at zero and imaginary chemical potential in 
Refs.~\cite{Braun:2006jd,Braun:2009gm}. With Dyson-Schwinger equations,
lattice data for the quenched gluon where used as an input for the quark DSE to 
study chiral and deconfinement transitions as well as quark spectral functions in 
quenched QCD \cite{Fischer:2009wc,Fischer:2010fx}. These calculations were generalized
to the two-flavor case and to finite real chemical potential in Ref.~\cite{arXiv:1104.1564}.

In the latter study the quarks have been back-coupled to the Yang-Mills sector using 
a hard thermal loop (HTL) like approximation for the quark loop in the gluon DSE.
In this work we will lift this restriction and consider the
back-reaction with dressed quarks, as discussed in section \ref{sec:DSE} and the appendix. We will 
therefore be able for the first time to calculate the unquenched two plus one 
flavor quark {\it and} gluon propagators at finite temperature and density.  
We also determine the chiral and deconfinement transition regions from the quark 
condensate and the dressed Polyakov loop 
\cite{Gattringer:2006ci,Synatschke:2007bz,Bilgici:2008qy,Fischer:2009wc}.
Our results for the QCD phase diagram with $N_f=2$ and $N_f=2+1$ quark flavors
are presented and discussed in section \ref{sec:results}. 
In the concluding section \ref{sec:sum} we summarize our results and discuss 
remaining uncertainties due to our truncation scheme.

\section{In-medium propagators and order parameters \label{sec:DSE}}

The objects of interest in this work are the propagators of the quark and the gluon,
from which we will extract their thermal properties and order parameters for chiral 
symmetry breaking and confinement. In the Landau gauge and in the medium they are 
given by

\begin{eqnarray}
S(p) &=& [i(\omega_n+i\mu)\gamma_4C(p)+i\vec{p}\vec{\gamma}A(p)+B(p)]^{-1}\,,\nonumber\\ \label{eq:qProp}\\
D_{\mu\nu}(p) &=& P_{\mu\nu}^L(p)\frac{Z^L(p)}{p^2} + P_{\mu\nu}^T(p)\frac{Z^T(p)}{p^2}\,.
\end{eqnarray}
The vector dressing functions $A$ and $C$ and the scalar dressing function $B$ of the 
quark propagator depend on the momentum $p=(\omega_n,\vec{p})$ and, implicitly, on 
temperature and chemical potential. The same is true for the gluon dressing functions
$Z_L$ and $Z_T$ with longitudinal and transversal orientation with respect
to the heat bath. The corresponding projectors are given by

\begin{eqnarray}
P_{\mu\nu}^T &=& \left(1-\delta_{\mu 4}\right)\left(1-\delta_{\nu 4}\right)\left(\delta_{\mu\nu}-\frac{p_\mu p_\nu}{\vec{p}^2}\right), \label{eq:projT} \\
P_{\mu\nu}^L &=& P_{\mu\nu} - P_{\mu\nu}^T. \label{eq:projL}
\end{eqnarray}
The Matsubara modes $\omega_n$ are subject to antiperiodic boundary conditions for fermions,
$\omega_n=\pi T(2n+1)$, and periodic ones for the gauge boson, $\omega_n=\pi T2n$.
The quark dressing-functions will be determined self-consistently from the quark DSE,
while the gluon dressing-functions will be given by a combination of lattice results
and the gluon DSE as discussed in section \ref{sec:gluonDSE} below.

From the quark propagator, Eq.~(\ref{eq:qProp}), one can calculate the condensate
\beq \label{eq:condensate}
\langle\bar{\psi}\psi\rangle = 
Z_2T\sum_n\int\frac{d^3p}{(2\pi)^3}\mathrm{Tr}_D\left[S(p)\right],
\eeq
where $Z_2$ is the quark wave function renormalization constant. For finite bare quark
masses the resulting quantity is quadratically divergent and can be regularized according to
\begin{equation}
\Delta_{l,s} = \langle\bar\psi\psi\rangle_l - \frac{m_l}{m_s}\langle\bar\psi\psi\rangle_s\,,
\label{eq:cond_renorm}
\end{equation}
where the divergent part $m\Lambda^2$ from the strange quark condensate 
$\langle\bar\psi\psi\rangle_s$ cancels the 
corresponding part of the light quark condensate $\langle\bar\psi\psi\rangle_l$ when
multiplied with the ratio $m_l/m_s$ of bare light to strange quark masses. The quark 
condensate is a strict order parameter for chiral symmetry breaking in the chiral limit 
and serves as an indicator for the chiral crossover at physical quark masses.

In \cite{arXiv:1104.1564} we presented a first calculation of the dressed Polyakov loop 
at finite real chemical potential in two-flavor QCD. This object is defined by
\cite{Gattringer:2006ci,Synatschke:2007bz,Bilgici:2008qy}

\begin{equation}
\Sigma_n = \int_0^{2\pi}\frac{d\varphi}{2\pi}e^{-i\varphi n} \langle\bar\psi\psi\rangle_\varphi,
\end{equation}
where $\langle\bar{\psi}\psi\rangle_\varphi$ 
is the quark condensate evaluated at generalized, $U(1)$-valued boundary conditions 
$\psi(\vec{x},1/T)=e^{i\varphi}\psi(\vec{x},0)$ with $\varphi \in [0,2\pi[$. 
For 
$\Sigma_{\pm 1}$ to act as order parameters for deconfinement it is mandatory to 
implement the generalized, $U(1)$-valued boundary conditions only on the level of 
observables, but not in the partition function itself. All closed quark loops 
therefore maintain the physical value $\varphi = \pi$, whereas $\varphi \in [0,2\pi[$ 
in the quark-DSE. This procedure breaks the Roberge-Weiss symmetry, a necessary 
condition for the dual condensate to act as an order parameter for center symmetry 
breaking \cite{Braun:2009gm,Fukushima:2003fm}.

In the presence of a crossover for chiral symmetry breaking and confinement, we determine
the maxima of the derivative with respect to the quark mass of the order parameters to obtain the
pseudo-critical temperatures.

\begin{figure}
\centering\includegraphics[width=0.4\textwidth]{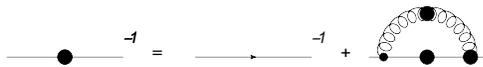}
\caption{The DSE for the quark propagator. Large blobs denote dressed 
propagators and vertices. \label{fig:quarkDSE}}
\end{figure}

The main difficulty in the Dyson-Schwinger framework is to find a truncation scheme
that correctly describes the relevant physics. In the case of QCD thermodynamics this
means to cut off the infinite tower of DSEs in such a way that the temperature and density
dependence of the $n$-point functions that are not determined self-consistently is carefully
approximated.
Fig.~\ref{fig:quarkDSE} shows the DSE for the quark propagator, which depends
on the fully dressed gluon propagator and quark-gluon vertex. For the quark-gluon vertex we will 
use a carefully designed expression constructed along its Slavnov-Taylor identity 
and matched to provide the correct renormalization group running in the quark and
gluon DSEs. For the gluon we will use temperature dependent lattice data for the
quenched propagator, and unquench it by invoking the quark part of the gluon DSE. 
In particular this means that the gluon becomes sensitive to the chiral dynamics 
of the quark sector.

\subsection{The gluon DSE \label{sec:gluonDSE}}

\begin{figure}[t!]
\centering\includegraphics[width=0.45\textwidth]{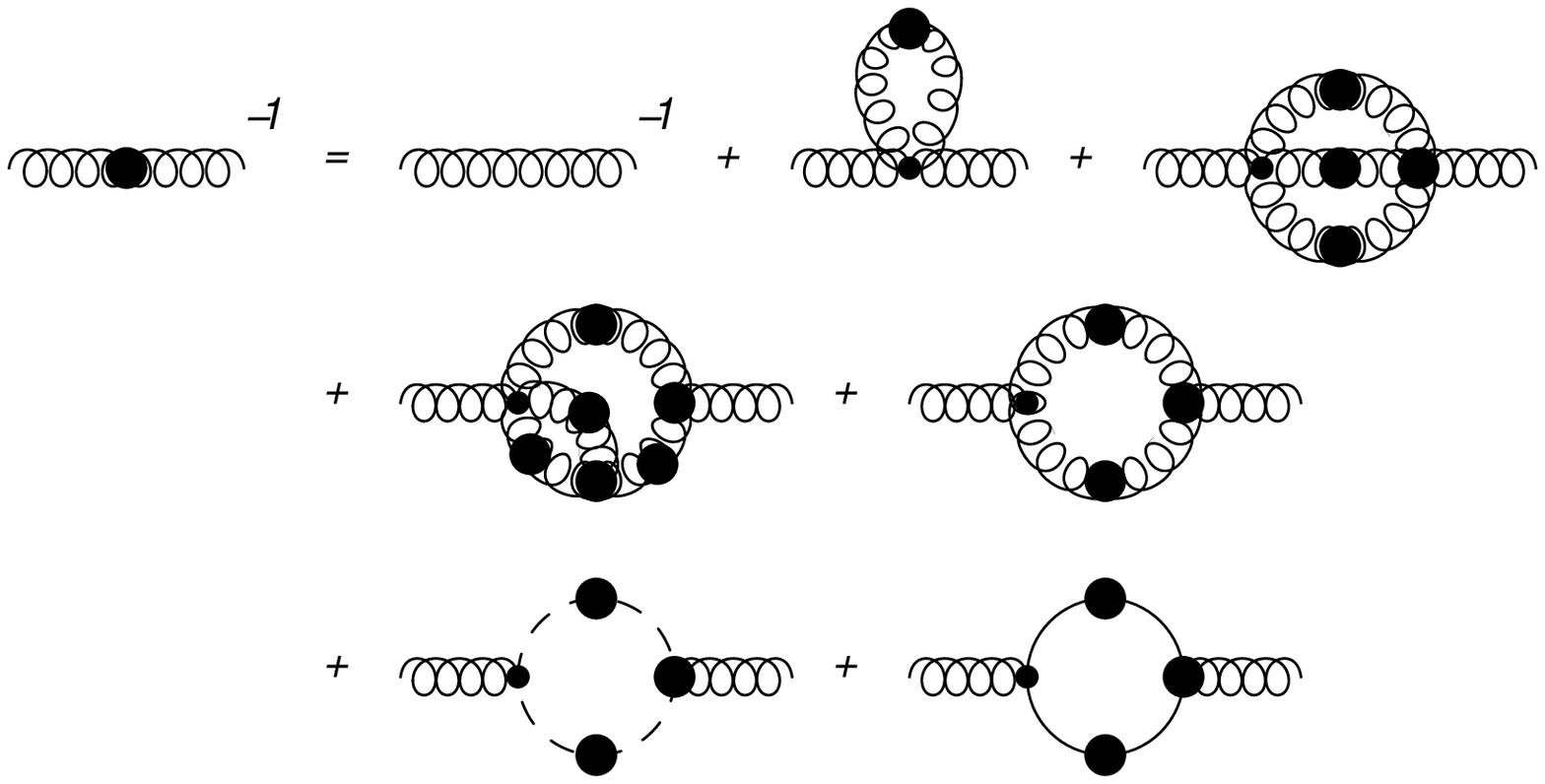}
\caption{The full gluon DSE. Blobs denote dressed propagators and vertices. 
\label{fig:fullGluonDSE}}\vspace*{3mm}
\includegraphics[width=0.4\textwidth]{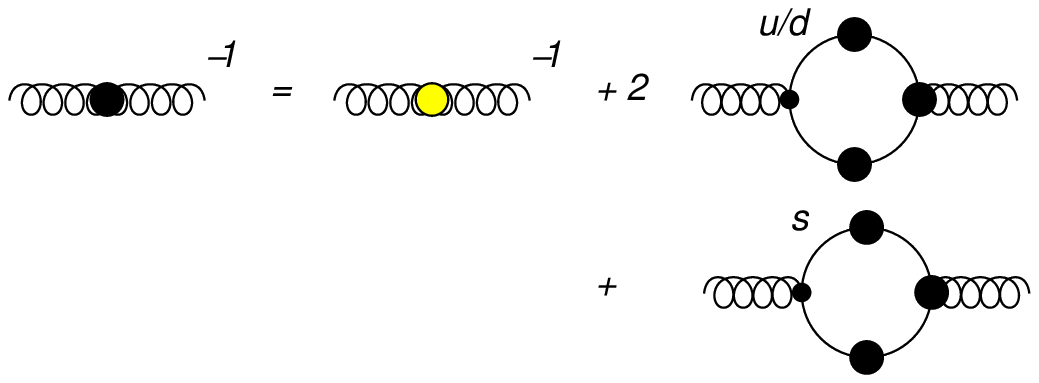}
\caption{The truncated gluon DSE for $N_f=2+1$ QCD. The yellow dot denotes the quenched propagator. 
(For interpretation of the references to color in this figure
legend, the reader is referred to the web version of this Letter.) \label{fig:apprGluonDSE}}
\end{figure}

Fig.~\ref{fig:fullGluonDSE} shows the untruncated gluon DSE. Studies of this equation at 
finite temperature have proven to be very difficult \cite{Cucchieri:2007ta,Maas:2005ym},
and are mostly restricted to large and infinite temperatures. Another possibility to access 
the finite-temperature gluon is to use the functional renormalization group, where 
interesting first results have been reported recently \cite{Fister:2011uw}. Although the
systematic errors are still considerable, probably the most reliable resource, however, is 
lattice QCD, where a number of quenched studies are available 
\cite{Cucchieri:2007ta,Fischer:2010fx,Aouane:2011fv,Maas:2011ez,Cucchieri:2012nx}. These
errors are most pronounced around the deconfinement transition temperature of the pure
gauge theory, but fortunately have not much impact on our unquenched 
calculation \cite{arXiv:1111.0180}. Here, we will continue to use the lattice data 
from Ref.~\cite{Fischer:2010fx} as also done in our previous work \cite{arXiv:1104.1564}.

In terms of diagrams, quenched QCD means that all quark loops are neglected,
{\it i.e.} the last diagram of Fig.~\ref{fig:fullGluonDSE}. One then recovers
the first (second) order deconfinement transition of SU(3) (SU(2)) Yang-Mills
theory at the correct transition temperatures. Using the dressed Polyakov loop 
this transition has been studied in 
Ref.~\cite{Fischer:2010fx}, a corresponding study for the Polyakov loop potential
using FRGs is reported in Ref.~\cite{Braun:2007bx}. In order to investigate the 
phase diagram at physical quark masses, however, one needs to take the quark loop 
explicitly into account. In fact, as we will see a careful treatment of the back-reaction
of the quarks onto the gluon sector is mandatory for a meaningful study of the phase
diagram. Consequently, this also introduces an implicit chemical-potential dependence 
of the gluon dressing functions which is most pronounced in their thermal masses, see
section \ref{2p1results}.

Technically, we implement this dependence by substituting the Yang-Mills part of 
the gluon DSE by the quenched propagator from the lattice, and merely add the quark loop.
The resulting equation is shown diagrammatically in Fig.~\ref{fig:apprGluonDSE},
where we already distinguished between the (isospin-symmetric) light quark and the 
strange quark contribution. This procedure of merely adding the quark loops does 
neglect all quark loop effects inside the gluon or ghost loops and is therefore
only approximate. However, we explicitly checked that this approximation works well 
in the vacuum \cite{Fischer:2003rp}, where differences to the fully self-consistent 
solution are below the five percent level. This may still be the case at 
finite temperature and chemical potential.

In \cite{arXiv:1104.1564} we furthermore approximated the quark loop by taking
only bare quarks into account, which amounts to using a hard-thermal loop (HTL) 
like approximation with only the quark-gluon vertex being dressed (for a review
on HTL results see e.g. \cite{Blaizot:2001nr}). This approximation,
introduced for technical reasons, is much more severe and can only be justified 
for temperatures well above the critical one. In this work we therefore improve
upon this situation and calculate the quark loop explicitly with the quark dressing
functions obtained from the quark-DSE. By doing so, the quark and gluon DSE become 
coupled and have to be solved simultaneously. This means also, that the gluon becomes 
sensitive to the chiral dynamics in the quark sector and in particular to the 
chiral transition. The coupled system of DSEs in Figs.~\ref{fig:quarkDSE} 
and Fig.~\ref{fig:apprGluonDSE} can be written as
\begin{widetext}
\begin{align*}
\left[S^{f}(p)\right]^{-1} &= Z^f_{2}\left[S_0^{f}(p)\right]^{-1} 
+ C_{F}Z^f_{2}Z^f_{1F} \is_l g\gamma_\mu S^f(l) g\Gamma^f_\nu(l^2,p^2,q^2) D_{\mu\nu}(q), \\
D_{\mu\nu}^{-1}(p) &= \left[D_{\mu\nu}^{qu.}(p)\right]^{-1} - \sum_{f}^{N_f}\frac{Z_2^f}{2}\is_l \Tr\left[ g\gamma_\mu S^{f}(l) g\Gamma^f_\nu(l^2,q^2,p^2) S^{f}(q)\right],
\end{align*}
\end{widetext}
where $q=p-l$, $S^f$ is the quark propagator for one flavor $f=u,d,s$, $C_F=\frac{4}{3}$ 
is the Casimir operator, $\Gamma_\nu$ the dressed quark-gluon vertex, $Z_{1F}$ 
and $Z_2$ are the vertex and wave function renormalization constants, and the 
Matsubara sum as well as the integration over the loop three-momentum $\vec{l}$ is 
represented by
$\is_{\;\;l} = T\sum_n \int\frac{d^3l}{(2\pi)^3}$. For the momenta we often abbreviate
generically $p = (\vec{p},\omega_p)$ and $p^2 = \vec{p}^2 + \omega_p^2$.
With the transversal and longitudinal projectors from Eqs.~(\ref{eq:projT},\ref{eq:projL}), the quark loop can be decomposed as
\begin{equation}
\Pi_{\mu\nu}(p) = P^T_{\mu\nu} \Pi^T(p) + P^L_{\mu\nu} \Pi^L(p).
\end{equation}

The medium effects in the gluon-DSE manifest themselves predominantly in contributions
to the thermal mass of the gluon. As a result, the quark loop can be split into a 
finite part that is similar to the vacuum quark loop and an infrared divergent thermal part 
that is proportional to $1/p^2$:
\begin{equation}
\Pi^{T,L}(\vec{p}^2,\omega_p) = \frac{\left(m^{T,L}_{th.}\right)^2}{2 p^2}
 + \Pi^{T,L}_{reg.}(\vec{p}^2,\omega_p), \label{mediumquark}
\end{equation}
with the infra-red regular part $\Pi_{reg.}$ and thermal masses
\begin{equation}
\left(m^{T,L}_{th.}\right)^2 = 2\left.\Pi^{T,L}(\vec{p}^2,\omega_p=0)\vec{p}^2\right|_{
\vec{p} \rightarrow 0}\,, \label{eq:mmth}
\end{equation}
A technical problem associated with the use of a sharp momentum cut-off $\Lambda$ in the 
quark loop is the appearance of quadratic divergencies. This problem is particularly 
troublesome in the medium, since the $\Lambda^2$ term appears like a squared thermal mass 
and has to be carefully removed without spoiling the $T^2$ and $\mu^2$ terms which carry the
physics. We do this using the Brown-Pennington projection method \cite{Brown:1988bm},
as detailed in Appendix~\ref{app:qlNumerics}.
After removing the quadratic divergences and projecting on longitudinal and transversal
parts the quark loop in the medium is given by

\begin{widetext}
\begin{align}
\Pi^T(\vec{p}^2,\omega_p=0) &= \frac{g^2}{2} \is_l \frac{\Gamma(l^2,q^2,p^2)}{D_q(l)D_q(q)}
\left\{
A(l)A(q)\Gamma_s(l,q) \left( 3\frac{\vec{l}\cdot\vec{p}^2}{\vec{p}^2} + 2\vec{l}\vec{p} - \vec{l}^2 \right)
\right\} \label{eq:PiT} \\
\Pi^L(\vec{p}^2,\omega_p=0) &= \frac{g^2}{2} \is_l \frac{\Gamma(l^2,q^2,p^2)}{D_q(l)D_q(q)}
\left\{
A(l)A(q)\left[\Gamma_s(l,q) \left( 2\frac{\vec{l}\cdot\vec{p}\vec{p}\cdot\vec{q}}{\vec{p}^2} - \vec{l}\vec{q} \right)+ \Gamma_4(l,q) \vec{l}\cdot\vec{q} \right]\right. \notag \\
&\;\left.\phantom{\frac{\vec{a}}{\vec{b}}}
+ B(l)B(q)\left[ \Gamma_4(l,q) - \Gamma_s(l,q) \right] + C(l)C(q)\left[ -\tilde\omega_n^2(\Gamma_s(l,q)+\Gamma_4(l,q))\right]
\right\} \label{eq:PiL}
\end{align}
\end{widetext}
with $q=p-l$ and $D_q(p)=\vec{p}^2A^2(p)+\tilde\omega_p^2C^2(p)+B^2(p)$
for one quark flavor and with the abbreviation $\tilde\omega_n=\omega_n+i\mu$.
The quark-gluon vertex has been split in three scalar functions, $\Gamma$, $\Gamma_s$ and $\Gamma_4$
which will be explained below. For the limit $\vec{p} \rightarrow 0$ we find
the thermal masses. Of course, the expressions (\ref{eq:mmth}) together 
with (\ref{eq:PiT}),(\ref{eq:PiL}) also have the correct high-temperature limit,
where the HTL results $\left(m^T_{th.}\right)^2=0$ and
$\left(m^L_{th.}\right)^2=\frac{g^2}{12}\left(T^2 + 3\frac{\mu^2}{\pi^2}\right)$,
for one flavor, are recovered.

Some technical details of the evaluation of this object are given in App.~\ref{app:qlNumerics}.
It is noteworthy that the quark loop does not contribute to the transversal thermal masses,
in contrast to the Yang-Mills part of the gluon self-energy.

\subsection{The quark-gluon vertex \label{sec:vertex}}

An important quantity that appears in both, the quark and gluon DSEs, is the dressed 
quark-gluon vertex. At zero temperature the nonperturbative structure of this vertex 
has been explored to some degree in the past \cite{Alkofer:2008tt,Fischer:2009jm,Chang:2010hb},
however not much is known about the behavior at finite temperature and density.
Therefore in order to close the system of DSEs we have to resort to an educated guess
for the form of the quark-gluon interaction. Fortunately, there are constraints which
serve as important guides in the construction of such a guess. On the one hand, at
large momenta temperature and density effects are exponentially suppressed and one
can entirely rely on the vacuum structure of the vertex. It is then well known, that 
the vertex has to combine with the gluon dressing functions such that the resulting combination
of dressing functions runs like the running coupling in the ultraviolet. In addition,
the vertex has to satisfy its Slavnov-Taylor identity. This identity has not been
solved yet in QCD, however important guidance can be obtained from the corresponding 
Abelian Ward identity, which serves to express part of the vertex dressing in terms of
the quark dressing functions. All these constraints have already been taken into
account in previous works \cite{Fischer:2009wc,Fischer:2010fx,arXiv:1104.1564}. 
We therefore employ the same construction which uses the first term of the Ball-Chiu vertex,
satisfying the Abelian Ward-Takahashi identity, multiplied with an infrared enhanced 
function $\Gamma(p^2,k^2,q^2)$ that accounts for the non-Abelian dressing effects
and the correct ultraviolet running of the vertex. The resulting expression reads
\begin{widetext}
\begin{eqnarray}
\Gamma_\mu(p,k;q) &=& \gamma_\mu\cdot\Gamma(p^2,k^2,q^2) \cdot 
\left(\delta_{\mu,4}\frac{C(p)+C(q)}{2} + \delta_{\mu,i}\frac{A(p)+A(q)}{2} \right), \nonumber\\
\Gamma(p^2,k^2,q^2) &=& \frac{d_1}{d_2+x} \!
 + \!\frac{x}{\Lambda^2+x}
\left(\frac{\beta_0 \alpha(\mu)\ln[x/\Lambda^2+1]}{4\pi}\right)^{2\delta} \label{vertex}
\end{eqnarray}
\end{widetext}
where $p$ and $k$ are the fermion momenta, $q$ is the gluon momentum
and $d_1$ as well as the scales $d_2$ and $\Lambda$ are parameters. Whereas
$d_2$ and $\Lambda$ control the renormalization group running of the vertex 
function from the large into the low momentum region, $d_1$ controls the
strength of the quark-gluon interaction at small momenta and therefore the
amount of quark mass generation in the hadronic phase.
In the ultra-violet $\delta=-9\frac{Nc}{44N_c - 8N_f}$ is the anomalous
dimension of the vertex and $\beta_0=\frac{11N_c-2N_f}{3}$.
In Eqs.~(\ref{eq:PiT},\ref{eq:PiL}) we have used the abbreviations
$\Gamma_s(p,q)=\frac{A(p)+A(q)}{2}$ and $\Gamma_4(p,q)=\frac{C(p)+C(q)}{2}$.

The squared momentum variable $x$ is chosen to be $q^2$ in the quark DSE, 
but $p^2+k^2$ in the quark loop. This change in the momentum dependence 
is well justified by the need to maintain multiplicative renormalizability
in the gluon-DSE \cite{Fischer:2003rp}. Note that, since the Ball-Chiu part 
is determined from the quark dressing functions our vertex does include 
effects from temperature and chemical potential.

\subsection{Strange quarks and unquenching effects}

The evaluation of the quark loop with dressed quarks allows us to add strange quarks
in a non-trivial way. In general the light and strange quarks couple via unquenching 
effects in QCD, as shown in Fig.~\ref{fig:apprGluonDSE}. In the vacuum, results for 
the propagators of $N_f=2+1$ QCD have been discussed in Ref.~\cite{Fischer:2003rp}. 
Here we present the first finite temperature and chemical potential study of the 
propagators in the unquenched theory. However, we also wish to emphasize that not
all the unquenching effects are included yet. In addition to the ones in the gluon 
propagator there are also corresponding effects in the quark-gluon vertex. 
Diagrammatically,§ part of these can be written as hadronic contributions like pion
and kaon exchange \cite{Fischer:2009jm}. In Ref.~\cite{Fischer:2011pk} it has been 
argued that these effects are necessary to obtain critical scaling beyond mean field 
at the chiral phase transition temperature in the chiral limit. In our case of physical 
quark masses, however, we expect their influence on the critical temperatures to be 
small. Furthermore, since mesons have zero quark number the chemical potential 
dependence of the meson exchange is expected to be small as well. In this work we
therefore do not include these contributions explicitly, but rather absorb their
effects in our vertex ansatz (\ref{vertex}). While this strategy may work rather well for 
meson contributions, it presumably fails to account for diquark and baryon contributions
at large densities and low temperatures. This is the region of the phase diagram where 
nuclear effects are expected to play a major role \cite{Weise:2012yv} and results in 
our present truncation scheme should be interpreted with great care. We will come back 
to this point below. 

In our calculation we only vary the chemical potential of the up/down quarks and set 
the strange quark chemical potential to $\mu_s = 0$, since the net number of strange 
quarks in heavy ion collisions vanishes.
The light quark mass can be fixed by calculating the pion mass via the 
Gell--Mann-Oakes-Renner relation and the Pagel-Stoker approximation for the pion 
decay constant. This gives a mass of $m_{u/d} \approx 2$ MeV at a renormalization 
point of $80$ GeV. We furthermore fix the strange quark mass by its ratio 
to the light quark mass by choosing $\frac{m_s}{m_{u/d}} = 27$.
For the vertex parameters we take $d_1 = 7.5 \,\mbox{GeV}^2$, $d_2 = 0.5 \,\mbox{GeV}^2$
and $\Lambda = 1.4\,\mbox{GeV}$. For the coupling constant we take $\alpha(\mu) = 0.3$.
With this set of parameters and with two massless and one strange quark, the pion decay
constant in the vacuum is $f_\pi \approx 88$ MeV and at the physical point of quark masses
at vanishing density, $T_c \approx 156$ MeV. Those numbers agree very well with expectations,
which is in contrast to simpler truncation schemes where usually the critical temperature
comes out rather small \cite{Blank:2010bz}.

Note again, however, that unquenching effects in the vertex are taken into 
account indirectly via the 
quark dressing functions $A$ and $C$ that appear explicitly in the vertex 
construction (\ref{vertex}). We emphasize that it is nontrivial that 
this vertex construction leads to correct quenched \cite{Fischer:2010fx} as well 
as unquenched transition temperatures as will be detailed in the next section.

\section{Results \label{sec:results}}

\subsection{Condensates, thermal gluon mass and phase diagram for $N_f=2$}

We first discuss the effect of coupling the two-flavor quark and gluon DSEs 
together as explained above. Here we will encounter some qualitative effects,
which are still valid in the $N_f=2+1$ case discussed below.
\begin{figure}
\includegraphics[width=0.49\textwidth]{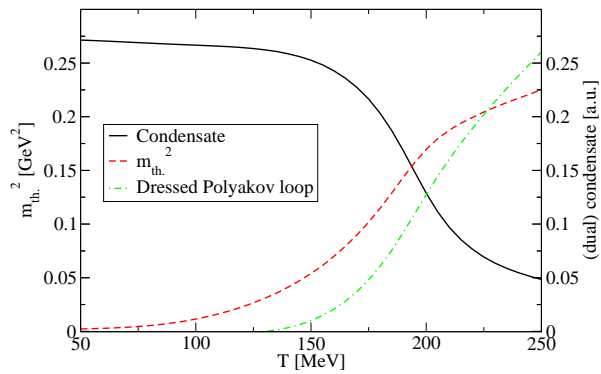}
\caption{Quark condensate, dressed Polyakov loop (dual condensate) and thermal electric gluon masses
for vanishing quark chemical potential with $N_f=2$.
\label{fig:thermalMassMu0}}
\end{figure}

The quark condensate and dressed Polyakov loop which are shown in 
Fig.~\ref{fig:thermalMassMu0} for two light quark flavors at vanishing chemical 
potential exhibit a crossover behavior around $T_c \approx 200$ MeV, with the 
transition region for chiral symmetry restoration and deconfinement being almost 
equal. We clearly identify the two most important effects from including the 
quark loop: a reduction of $T_c$ and a change from a first order phase transition to a crossover.
Fig.~\ref{fig:thermalMassMu0} also shows the contribution from the quark loop to the 
electric part of the thermal gluon-mass as defined in Eqs.~(\ref{eq:mmth}),(\ref{eq:PiL}). 
The thermal mass is small in the hadronic 
phase, shows a sharp rise in a temperature range around the transition temperature 
and converges to the $T^2$-behavior expected from HTL-calculations in the high
temperature phase. This behavior can be expected, since the quark loop has an 
inverse dependence on the quark mass-function. It is therefore suppressed by 
dynamical chiral symmetry breaking in the hadronic phase and becomes large in 
the high temperature phase, where the quark condensate becomes small. 
As compared to a truncation scheme where the quarks in 
the loop are not dressed, {\it i.e.} the HTL-like approximation we used 
in \cite{arXiv:1104.1564}, the quark condensate shows a steeper transition
around $T_c$. This behavior originates from the growing thermal gluon mass 
due to the quark loop, which in turn decreases the interaction strength in 
the quark DSE. Therefore, the back coupling of the quark and gluon DSEs 
accelerates the chiral phase transition.

\begin{figure}
\includegraphics[width=0.49\textwidth]{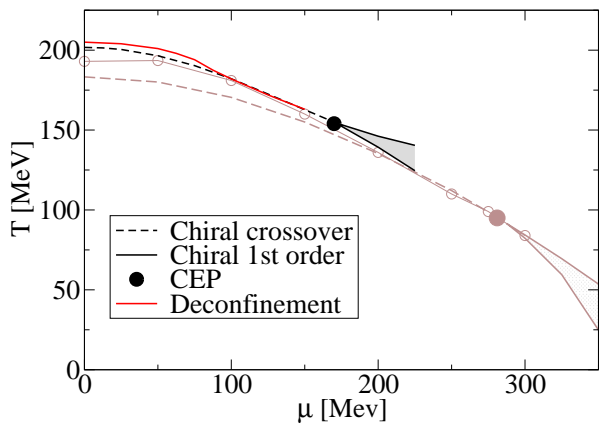}
\caption{The phase diagram for $N_f=2$ in the temperature and (up/down) quark chemical 
potential plane. Shown are the transition lines determined
using the improved quark loop of this work (see legend) compared with our previous results
(light colors in background)
in the HTL-approximation of the quark loop \cite{arXiv:1104.1564}. \label{fig:phaseDiagramNf2}}
\end{figure}
In Fig.~\ref{fig:phaseDiagramNf2} we present our results for the chiral and
deconfinement transition lines determined in this work and compared to our previous
results of Ref.~\cite{arXiv:1104.1564}, where we used the HTL approximation for the
quark loop. Note that the deconfinement transition is determined solely from
$\Sigma_{-1}$; since the corresponding calculation of $\Sigma_{+1}$ is plagued by
considerable numerical uncertainties we refrain from showing the results.
We first observe that the pseudo-critical 
temperatures for restoration of chiral symmetry and deconfinement are now closer
together, in comparison to the HTL results where a difference of up to $10$ MeV was 
observed. This closer matching of the critical temperatures can be explained by 
the gluon becoming sensitive to the chiral phase transition due to the explicit
appearance of the quark mass functions in the full quark loop calculation.
Moreover, the critical end-point has moved 
to a smaller quark chemical potential and a larger temperature. We attribute this 
to the accelerated chiral transition due to the improved back-coupling of the
quarks onto the gluon, as explained above. The effect is substantial and underlines
once more the general importance of details in the back-coupling effects from the 
quarks onto the Yang-Mills sector of QCD. As a result we obtain a ratio of critical 
quark chemical potential to critical temperature at the end point of 
$\mu^q_{EP}/T_{EP} \approx 1.1$. Thus for $N_f=2$ QCD this ratio stays above one 
(although not much) in agreement with the claim made in Ref.~\cite{arXiv:1104.1564}. 
We will see below that the addition of the strange quark further increases this ratio.

\subsection{Condensates and phase diagram of $N_f=2+1$ QCD \label{2p1results}}

\begin{figure}
\includegraphics[width=0.45\textwidth]{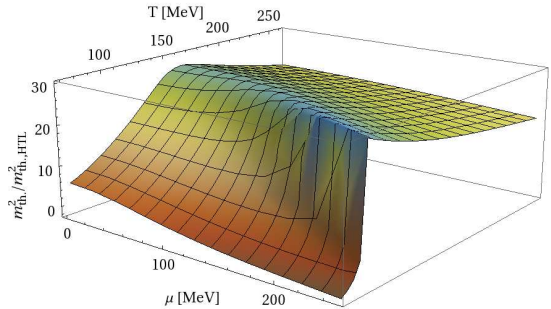}
\caption{The thermal gluon mass normalized by its asymptotic (HTL) value
(omitting $g^2$)
as a function of temperature and chemical potential. \label{fig:gluonmass}}
\end{figure}

In addition to the two light quark flavors we now also add the heavier strange
quark to the system. Recall, however, that following the conditions in a heavy ion 
collision we do not include a chemical potential for the strange quark in our
calculations. We discuss the effect of this choice below. Let us first have a 
look at the resulting unquenched gluon propagator. The main effect of the 
temperature and chemical potential dependent quark loop onto the gluon is the
change of the thermal electric mass. We plotted this quantity, normalized by
its asymptotic value, as a function of temperature and up/down-quark chemical
potential in Fig.~\ref{fig:gluonmass}.
One clearly sees the continuous change for small chemical potential similar
to the two-flavor case in Fig.~\ref{fig:thermalMassMu0}. For larger chemical 
potential the transition becomes steeper until it becomes discontinuous in 
the vicinity of the critical endpoint of the chiral transition, discussed 
below. This behavior underlines again the fact that the chiral transition
in the quark sector of the unquenched theory also drives the changes in the 
Yang-Mills sector of QCD. In the asymptotic region at large temperature and/or 
chemical potential we finally recover the HTL-result $m^2_{th} \sim T^2 + 3 \mu^2/\pi^2$
for the thermal gluon mass.

\begin{figure}
\includegraphics[width=0.45\textwidth]{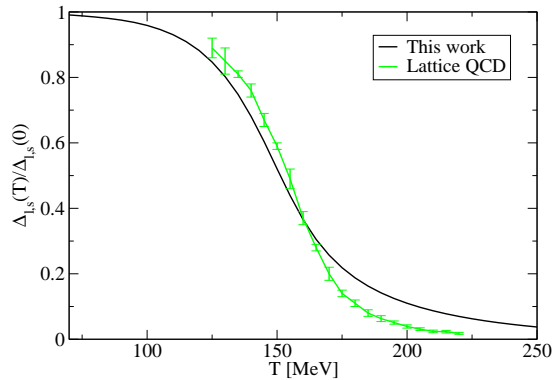}
\caption{The regularized condensate compared to lattice results from \cite{Borsanyi:2010bp}. \label{fig:DeltaVSlattice}}
\end{figure}

Let us now focus on the quark sector of QCD and come back to the behavior of 
the quark condensate at zero chemical potential. Since we have two plus one quark 
flavors at our disposal we can directly compare with corresponding lattice data 
at vanishing chemical potential. In Fig.~\ref{fig:DeltaVSlattice}, we compare 
the renormalized quark condensate $\Delta_{l,s}$ defined in Eq.~(\ref{eq:cond_renorm}) 
with lattice data from Ref.~\cite{Borsanyi:2010bp}.
The agreement is good except for the region above the phase transition,
where the DSE results are larger than the lattice data.
We attribute this discrepancy to the truncation of our quark-gluon vertex, which employs
medium effects only in the Ball-Chiu part, while the {\it ansatz} function $\Gamma$
is temperature independent.
This is in contrast to \cite{Mueller:2010ah}, where it has been suggested that the parameter $d_1$ of the
vertex infra-red part is reduced above $T_c$.
Therefore, neglecting the $T$-dependence of the vertex infra-red part leads to an over-estimation
of the interaction strength in the chirally symmetric phase, and thus to too large condensates.
Also the scalar parts in the vertex are neglected, which are known
to be important when it comes to the details of dynamical chiral symmetry breaking and
restoration \cite{Alkofer:2008tt}. Additionally, the (neglected) temperature dependence of 
explicit hadronic parts of the vertex may play a role \cite{Fischer:2011pk}.
Nevertheless, we find our present truncation sufficiently accurate in particular with
respect to the steepness of the crossover to justify its
extension into the finite chemical potential domain.
 
\begin{figure}
\includegraphics[width=0.45\textwidth]{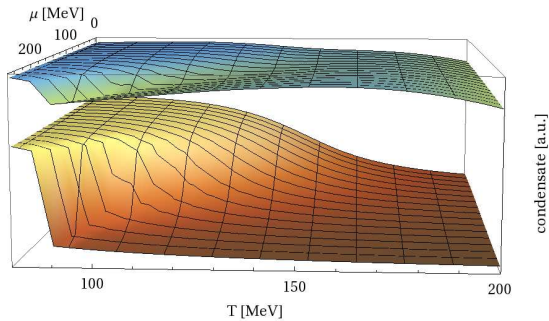}
\caption{The light (lower surface) and strange (upper surface) quark condensate as a function of temperature and 
chemical potential. \label{fig:Nf3condensates}}
\end{figure}
In Fig.~\ref{fig:Nf3condensates} we show our results for the light and strange quark 
condensates in the temperature and chemical-potential plane. The condensates are not
renormalized and shown on different scales for the sake of clarity.
In the background of the figure we see again the crossover behavior of the light quark condensate at small values of the
chemical potential. This turns into a first order transition at large chemical potential,
visible in the foreground of the figure. The strange quark condensate shows a behavior 
similar to the light quark condensate around the phase transition, but continues to melt 
at larger temperatures. Certainly, this behavior is caused by the larger explicit breaking
of chiral symmetry for the strange quark. At densities where the light quark shows a first 
order phase transition, the strange quark also shows a discontinuity at the same 
temperature. This is inherited from the discontinuity in the gluon propagator, see 
Fig.~\ref{fig:gluonmass}, and demonstrates how the gluon couples light and strange quarks.

\begin{figure}
\includegraphics[width=0.49\textwidth]{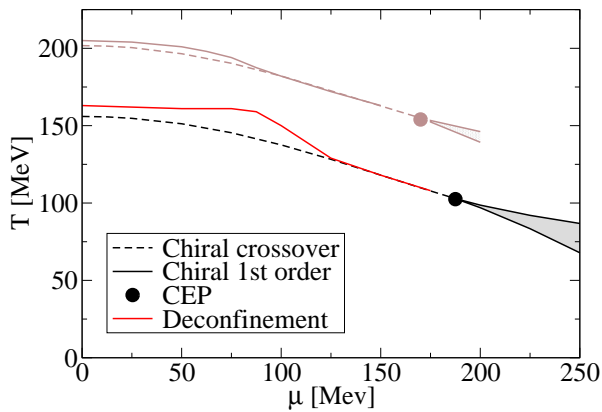}
\caption{The phase diagram for two plus one flavors.
  The light colors (top lines) show the $N_f=2$ results as a comparison. \label{fig:phaseDiagramNf2p1}}
\end{figure}

In Fig.~\ref{fig:phaseDiagramNf2p1} we show the resulting phase diagram for $2+1$ 
flavor QCD in our approximation scheme. 
As compared to the two-flavor case, Fig.~\ref{fig:phaseDiagramNf2}, we find a chiral
critical endpoint at slightly larger values of chemical potential. Thus, although the
additional strange quark lowers the transition temperatures by roughly $45$ MeV, the
melting of the condensate is slightly less steep as for the $N_f=2$ case. As a consequence,
one hits the first order transition region at larger values of the chemical potential.

At $\mu=0$ we furthermore observe a slightly wider spread between the chiral and 
the deconfinement transition for the $N_f=2+1$ case as compared to $N_f=2$. This 
effect may be explained by the different strength of the back-reaction of the quarks 
onto the gluon sector. For $N_f=2+1$ the back-reaction is stronger, resulting in
a wider crossover transition region which in turn allows for a wider spread between
the transition temperatures. On the other hand, the quenched calculation with no 
back-reaction is entirely driven by deconfinement and both transitions happen at 
the same temperatures \cite{Fischer:2010fx}.
      
So far, we set the strange quark chemical potential to zero. Setting it equal to the light
quark chemical potential has very little effect on the results.

From the chiral transition line one can extract the curvature $\kappa$, 
which is defined by
\begin{equation}
T_c(\mu) = T_c(0)\left[1 - \kappa \left(\frac{\mu}{T_c(0)}\right)^2\right], \label{eq:kappa}
\end{equation}
as the coefficient of a power expansion around $\mu=0$. This quantity can be 
measured on the lattice despite the fermion sign problem, and therefore 
serves as a comparison of the chemical potential dependence of the transition 
line. Two different lattice groups determined the curvature recently and
found $\kappa \approx 0.059(2)(4)$ \cite{Kaczmarek:2011zz} and 
$\kappa \approx 0.059(20)$ \cite{Endrodi:2011gv} for $2+1$ flavors.
In comparison with the $N_f=2$ result $\kappa \approx 0.051(3)$ of 
Ref.~\cite{de Forcrand:2002ci}, there is a general trend of increasing 
curvature when more flavors are taken into account in agreement with 
a large $N_c$-analysis \cite{Toublan:2005rq}. In our case, 
we extract $\kappa$ by fitting Eq.~(\ref{eq:kappa}) to the chiral 
transition line in the region $\mu \in [0,25]$ MeV and $\mu \in [0,50]$ MeV 
for the case of the HTL-like quark loop. We also checked that the order 
$(\mu/T)^4$ is negligible in these intervals. 

Our results for the curvature $\kappa$ from the HTL-truncation of 
Ref.~\cite{arXiv:1104.1564} and the two results of this work
are summarized in Tab.~\ref{tab:CEPkappa} together with our findings for 
the location of the critical end-point (CEP). We find that the back-coupling 
increases the curvature compared to the HTL-like truncation. This is easily
understood from the shape of the transition lines of Fig.~\ref{fig:phaseDiagramNf2},
where the CEP from the fully back-coupled results is almost on top of the 
transition line of the HTL-approximated result, while at the same time the
transition temperature at $\mu=0$ is significantly larger. In general, our 
values for $\kappa$ agree in size with QM and PQM calculations 
\cite{Schaefer:2004en,Braun:2011iz} but are much larger than the 
lattice results. Part of this discrepancy may be attributed to volume effects 
in the lattice calculations \cite{Braun:2011iz}, another part of it may be
due to limitations of our present approximation scheme. This is particularly 
apparent in the $N_f$-dependence of the curvature. As discussed above, the
lattice results indicate a significant increase of curvature with $N_f$, 
whereas our results indicate the opposite. The reason for this discrepancy 
may be found in the incomplete treatment of unquenching effects in our 
present approach. While we took into account quark loop contributions in the 
gluon polarization, the inclusion of corresponding effects in the quark-gluon
vertex is numerically demanding and will be investigated in future work. 
\begin{table}
  \begin{tabular}{ | l || c | c |c| }
  \hline
    $N_f$	  & CEP		& $T_c(\mu=0)$  & $\kappa$	\\ \hline
    2 (HTL) & (280,90)	&183 MeV& 0.23	\\
    2	  & (171,154)	&202 MeV& 0.41 \\
    2+1	  & (190,100)	&156 MeV& 0.30 \\
  \hline
  \end{tabular}
\caption{Location of CEP and the curvature for $N_f=2$ and $N_f=2+1$ 
flavors as well as the $N_f=2$ flavor result in the HTL approximation 
of Ref.~\cite{arXiv:1104.1564}. \label{tab:CEPkappa}}
\end{table}

\section{Conclusion and outlook \label{sec:sum}}

In this work we have determined the phase boundaries of $N_f=2+1$ QCD
in the temperature and (up/down) quark chemical potential plane. We have
solved a coupled set of Dyson-Schwinger equations for the Landau gauge 
quark and gluon propagators. For the gluon we worked with lattice results 
for the temperature dependence of the electric and magnetic part of the
quenched propagator and added quark loop effects determined in the DSE-framework.
For the quark-gluon interaction we worked with a construction which
takes into account the correct renormalization group running of the vertex 
combined with leading order temperature and chemical potential effects 
as encoded in its Slavnov-Taylor identity. 

From the temperature and chemical potential dependence of the quark propagator
we extracted information on the chiral as well as the deconfinement transition
encoded in the quark condensate and the dressed Polyakov loop. As a result we 
find a crossover behavior at small chemical potential in quantitative
agreement with corresponding lattice results. We also find a critical endpoint
at $(T^{EP},\mu_q^{EP}) \approx (130,145)$ MeV followed by a first order 
transition. We therefore confirm our
previous claim of the nonexistence of a CEP for $\mu_q^{EP}/T^{EP} < 1$ 
\cite{arXiv:1104.1564} in agreement with extrapolated results of lattice 
QCD \cite{deForcrand:2010he,Endrodi:2011gv}.

The calculations presented in this work are, to our knowledge, the first
effort based on continuum QCD to determine the phase diagram with realistic 
up/down and strange quark masses combined with a dynamical gluon sector. 
In this respect we wish to emphasize again 
the importance of the back-coupling effects of the quarks onto the Yang-Mills 
sector of the theory. These effects are responsible for the change of the order of 
the transition from first order in the quenched theory to the crossover 
behavior of full QCD and the corresponding reduction of the (pseudo-)critical 
temperatures. Both phenomena are nicely reproduced in our present truncation scheme.

Nevertheless, this is not the end of the story and from the present approximation
scheme no firm conclusions on the precise location of the CEP nor on its very
existence can be drawn. The reason is to be found in our present expression
for the quark-gluon interaction. Although it does take into account the 
above-mentioned constraints from QCD itself, it lacks in detail when it comes 
to the precise dependence on temperature and chemical potential encoded
in hadron back-reaction effects onto the quark propagator, as discussed
in section \ref{sec:vertex}. In the DSE-framework,
these have only just begun to be explored \cite{Fischer:2011pk}. In particular,
baryon effects may be important when it comes to large chemical potential, 
as has been emphasized in \cite{Weise:2012yv} and explicitly checked in a
two-color framework in Ref.~\cite{Strodthoff:2011tz}. Furthermore, at large
chemical potential inhomogeneous phases may be energetically favored over 
homogeneous ones, see e.g. \cite{Kojo:2011cn,Carignano:2010ac} and references 
therein. Whether these effects show up in addition to the existence of a 
CEP or instead of the CEP is an important open question that needs to be 
investigated further in the future.

\section{Acknowledgements}

We thank Michael Buballa, Jens Braun, Leo Fister, Daniel M\"{u}ller, 
Jan Pawlowski, Bernd-Jochen Schaefer, 
Lorenz von Smekal and Jochen Wambach for fruitful discussions. This work has been 
supported by the Helmholtz Young Investigator Grant VH-NG-332 and the Helmholtz 
International Center for FAIR within the LOEWE program of the State of Hesse.

\appendix
\section{Details of the quark loop calculation \label{app:qlNumerics}}

\subsection{Vacuum}
The renormalization of the quark loop that is necessary to remove the logarithmic divergence is done in the vacuum,
since medium effects do not lead to new divergences.
We require that at the renormalization point $\zeta$, the self energy vanishes:

\begin{equation}
\Pi_{ren}(p^2) = \Pi(p^2) - \Pi(\zeta^2),
\end{equation}
where we choose $\zeta = 10$ GeV.

When solving the system of coupled DSEs in the vacuum, one notes that with the normal
kind of vertex {\it ansatz} which only depends on the gluon momentum,
as it is used in rainbow-ladder truncation, a function which allows for
dynamical chiral symmetry breaking is hard to find.
The reason for this is that the quark loop becomes very strong, as it is
multiplied with the vertex dressing function which only depends on the external momentum here.
Using a vertex which depends on all momenta, which a realistic vertex
certainly would, results in a weaker quark loop and therefore allows for chiral symmetry breaking.
Also, multiplicative renormalizability is spoiled by a dependence of the vertex
on the external momentum.

\subsection{Medium}
We already mentioned above that the dressing functions in the gluon DSE are plagued by
spurious quadratic divergences which appear when a hard cut-off is used.
In the vacuum they can be subtracted in several ways, in the medium one has
to be careful since they appear like thermal masses:

\begin{equation}
\Pi_L(0) = a \cdot \Lambda^2 + b \cdot T^2 + c \cdot \mu^2,
\end{equation}
and one has to be careful to only remove the first term.
We solve this problem by using a Brown-Pennington projection
\begin{equation}
\Pi_{\mu\nu}(p)^{(reg.)} = \Pi_{\mu\nu}(p) - \delta_{\mu\nu}\frac{p_{\alpha}p_{\beta}}{p^2}\Pi_{\alpha\beta}(p),
\end{equation}
which has been introduced in \cite{Brown:1988bm}.
With the Brown-Pennington projector it is vital to use an $O(4)$-invariant cut-off,
or the quadratic divergence appears again.
We achieve this by performing the Matsubara sum and the integral as follows:

\begin{equation}
T\sum_{n}\int d^3p \rightarrow T\sum_{n=-N-1}^{N}\int_{\epsilon^2}^{\Lambda'^2-\omega_n^2}d\vec{p}^2\frac{|\vec{p}|}{2}\int d\Omega,
\end{equation}
where $\Lambda' = \pi T(2N+2)$ is a temperature-dependent cut-off, that is chosen
close to the fixed cut-off $\Lambda$ by taking $N$ from the Matsubara mode closest to $\Lambda$.


\begin{thebibliography}{99}

\bibitem{Borsanyi:2010bp}
  S.~Borsanyi {\it et al.}  [Wuppertal-Budapest Collaboration],
  JHEP {\bf 1009} (2010) 073
  [arXiv:1005.3508 [hep-lat]].

\bibitem{Bazavov:2011nk}
  A.~Bazavov, T.~Bhattacharya, M.~Cheng, C.~DeTar, H.~T.~Ding, S.~Gottlieb, R.~Gupta and P.~Hegde {\it et al.},
  arXiv:1111.1710 [hep-lat].

\bibitem{latticemu}
  Z.~Fodor and S.~D.~Katz,
  JHEP {\bf 0203}, 014 (2002);
  [arXiv:hep-lat/0106002];
  S.~Ejiri, {\it et al. } 
  Prog.\ Theor.\ Phys.\ Suppl.\  {\bf 153}, 118 (2004);
  [arXiv:hep-lat/0312006];
  R.~V.~Gavai and S.~Gupta,
  Phys.\ Rev.\  D {\bf 71}, 114014 (2005);
  [arXiv:hep-lat/0412035];

\bibitem{deForcrand:2010he}
  P.~de Forcrand and O.~Philipsen,
  Phys.\ Rev.\ Lett.\  {\bf 105} (2010) 152001
  [arXiv:1004.3144 [hep-lat]].
  P.~de Forcrand and O.~Philipsen,
  PoS {\bf LATTICE2008} (2008) 208.
  [arXiv:0811.3858 [hep-lat]].

\bibitem{Kaczmarek:2011zz}
  O.~Kaczmarek, F.~Karsch, E.~Laermann, C.~Miao, S.~Mukherjee, P.~Petreczky, C.~Schmidt and W.~Soeldner {\it et al.},
  Phys.\ Rev.\ D {\bf 83} (2011) 014504
  [arXiv:1011.3130 [hep-lat]].

\bibitem{Endrodi:2011gv}
  G.~Endrodi, Z.~Fodor, S.~D.~Katz and K.~K.~Szabo,
  JHEP {\bf 1104} (2011) 001
  [arXiv:1102.1356 [hep-lat]].

\bibitem{Cea:2012ev}
  P.~Cea, L.~Cosmai, M.~D'Elia, A.~Papa and F.~Sanfilippo,
  arXiv:1202.5700 [hep-lat].

\bibitem{Weise:2012yv}
  W.~Weise,
  arXiv:1201.0950 [nucl-th].

\bibitem{Fukushima:2011jc}
  K.~Fukushima,
  J.\ Phys.\ G G {\bf 39} (2012) 013101
  [arXiv:1108.2939 [hep-ph]].

\bibitem{Fischer:2009tn}
  C.~S.~Fischer and J.~M.~Pawlowski,
  Phys.\ Rev.\ D {\bf 80} (2009) 025023
  [arXiv:0903.2193 [hep-th]];
  Phys.\ Rev.\ D {\bf 75} (2007) 025012
  [hep-th/0609009].

\bibitem{NJL}
  S.~P.~Klevansky,
  Rev.\ Mod.\ Phys.\  {\bf 64} (1992) 649,
  F.~Xu, H.~Mao, T.~K.~Mukherjee and M.~Huang,
  Phys.\ Rev.\ D {\bf 84} (2011) 074009
  [arXiv:1104.0873 [hep-ph]].

\bibitem{Fukushima:2003fw}
  K.~Fukushima,
  Phys.\ Lett.\  {\bf B591 } (2004)  277-284.
  [hep-ph/0310121].

\bibitem{Megias:2004hj}
  E.~Megias, E.~Ruiz Arriola and L.~L.~Salcedo,
  Phys.\ Rev.\ D {\bf 74} (2006) 065005
  [hep-ph/0412308].

\bibitem{Ratti:2005jh}
  C.~Ratti, M.~A.~Thaler, W.~Weise,
  Phys.\ Rev.\  {\bf D73 } (2006)  014019.
  [hep-ph/0506234].

\bibitem{Schaefer:2007pw}
  B.~-J.~Schaefer, J.~M.~Pawlowski, J.~Wambach,
  Phys.\ Rev.\  {\bf D76 } (2007)  074023.
  [arXiv:0704.3234 [hep-ph]].

\bibitem{Skokov:2010wb}
  V.~Skokov, B.~Stokic, B.~Friman and K.~Redlich,
  Phys.\ Rev.\ C {\bf 82} (2010) 015206
  [arXiv:1004.2665 [hep-ph]];
  V.~Skokov, B.~Friman and K.~Redlich,
  Phys.\ Rev.\ C {\bf 83} (2011) 054904
  [arXiv:1008.4570 [hep-ph]].

\bibitem{Herbst:2010rf}
  T.~K.~Herbst, J.~M.~Pawlowski, B.~-J.~Schaefer,
  Phys.\ Lett.\  {\bf B696 } (2011)  58-67.
  [arXiv:1008.0081 [hep-ph]].

\bibitem{Fister:2011uw}
  L.~Fister and J.~M.~Pawlowski,
  arXiv:1112.5440 [hep-ph].

\bibitem{Braun:2006jd}
  J.~Braun and H.~Gies,
  JHEP {\bf 0606} (2006) 024;
  J.~Braun,
  Eur.\ Phys.\ J.\  C {\bf 64} (2009) 459.

\bibitem{Braun:2009gm}
  J.~Braun, L.~M.~Haas, F.~Marhauser and J.~M.~Pawlowski,
  Phys.\ Rev.\ Lett.\  {\bf 106} (2011) 022002
  [arXiv:0908.0008 [hep-ph]].

\bibitem{Fischer:2009wc}
  C.~S.~Fischer,
  Phys.\ Rev.\ Lett.\  {\bf 103 } (2009)  052003;
  [arXiv:0904.2700 [hep-ph]].
  C.~S.~Fischer, J.~A.~Mueller,
  Phys.\ Rev.\  {\bf D80 } (2009)  074029.
  [arXiv:0908.0007 [hep-ph]].

\bibitem{Fischer:2010fx} 
  C.~S.~Fischer, A.~Maas and J.~A.~Muller,
  Eur.\ Phys.\ J.\ C {\bf 68}, 165 (2010)
  [arXiv:1003.1960 [hep-ph]].

\bibitem{arXiv:1104.1564}
  C.~S.~Fischer, J.~Luecker and J.~A.~Mueller,
  Phys.\ Lett.\ B\ {\bf 702} (2011) 438
  [arXiv:1104.1564 [hep-ph]].

\bibitem{Gattringer:2006ci}
  C.~Gattringer,
  Phys.\ Rev.\ Lett.\  {\bf 97} (2006) 032003.

\bibitem{Synatschke:2007bz}
  F.~Synatschke, A.~Wipf and C.~Wozar,
  Phys.\ Rev.\  D {\bf 75} (2007) 114003;

\bibitem{Bilgici:2008qy}
  E.~Bilgici, F.~Bruckmann, C.~Gattringer and C.~Hagen,
  Phys.\ Rev.\  D {\bf 77} (2008) 094007.

\bibitem{Fukushima:2003fm}
  K.~Fukushima,
  Phys.\ Rev.\  D {\bf 68} (2003) 045004.
  [arXiv:hep-ph/0303225].


\bibitem{Cucchieri:2007ta}
  A.~Cucchieri, A.~Maas and T.~Mendes,
  Phys.\ Rev.\ D {\bf 75} (2007) 076003
  [arXiv:hep-lat/0702022].

\bibitem{Maas:2005ym}
  A.~Maas,
  Mod.\ Phys.\ Lett.\ A {\bf 20} (2005) 1797
  [arXiv:hep-ph/0506066];
  A.~Maas, J.~Wambach and R.~Alkofer,
  Eur.\ Phys.\ J.\ C {\bf 42} (2005) 93
  [arXiv:hep-ph/0504019].

\bibitem{Aouane:2011fv}
  R.~Aouane, V.~G.~Bornyakov, E.~M.~Ilgenfritz, V.~K.~Mitrjushkin, M.~Muller-Preussker and A.~Sternbeck,
  Phys.\ Rev.\ D {\bf 85} (2012) 034501
  [arXiv:1108.1735 [hep-lat]].

\bibitem{Maas:2011ez}
  A.~Maas, J.~M.~Pawlowski, L.~von Smekal and D.~Spielmann,
  Phys.\ Rev.\ D {\bf 85} (2012) 034037
  [arXiv:1110.6340 [hep-lat]].

\bibitem{Cucchieri:2012nx}
  A.~Cucchieri and T.~Mendes,
  arXiv:1201.6086 [hep-lat].

\bibitem{arXiv:1111.0180}
  J.~Luecker and C.~S.~Fischer,
  arXiv:1111.0180 [hep-ph].

\bibitem{Braun:2007bx}
  J.~Braun, H.~Gies, J.~M.~Pawlowski,
  Phys.\ Lett.\  {\bf B684 } (2010)  262-267.

\bibitem{Fischer:2003rp}
  C.~S.~Fischer and R.~Alkofer,
  Phys.\ Rev.\ D {\bf 67} (2003) 094020
  [hep-ph/0301094];
  C.~S.~Fischer,
  J.\ Phys.\ G G {\bf 32} (2006) R253
  [hep-ph/0605173].
  
\bibitem{Blaizot:2001nr}
  J.~-P.~Blaizot and E.~Iancu,
  Phys.\ Rept.\  {\bf 359} (2002) 355
  [hep-ph/0101103].

\bibitem{Brown:1988bm}
  N.~Brown and M.~R.~Pennington,
  Phys.\ Rev.\ D {\bf 38} (1988) 2266.

\bibitem{Alkofer:2008tt}
  R.~Alkofer, C.~S.~Fischer, F.~J.~Llanes-Estrada and K.~Schwenzer,
  Annals Phys.\  {\bf 324} (2009) 106
  [arXiv:0804.3042 [hep-ph]].

\bibitem{Fischer:2009jm}
  C.~S.~Fischer and R.~Williams,
  Phys.\ Rev.\ Lett.\  {\bf 103} (2009) 122001
  [arXiv:0905.2291 [hep-ph]];
  Phys.\ Rev.\ D {\bf 78} (2008) 074006
  [arXiv:0808.3372 [hep-ph]].

\bibitem{Chang:2010hb}
  L.~Chang, Y.~-X.~Liu and C.~D.~Roberts,
  Phys.\ Rev.\ Lett.\  {\bf 106} (2011) 072001
  [arXiv:1009.3458 [nucl-th]].

\bibitem{Fischer:2011pk}
  C.~S.~Fischer and J.~A.~Mueller,
  Phys.\ Rev.\ D {\bf 84} (2011) 054013
  [arXiv:1106.2700 [hep-ph]].

\bibitem{Blank:2010bz}
  M.~Blank and A.~Krassnigg,
  Phys.\ Rev.\ D {\bf 82} (2010) 034006
  [arXiv:1004.5301 [hep-ph]].

\bibitem{Mueller:2010ah}
  J.~A.~Mueller, C.~S.~Fischer and D.~Nickel,
  Eur.\ Phys.\ J.\ C {\bf 70} (2010) 1037
  [arXiv:1009.3762 [hep-ph]].
 
\bibitem{de Forcrand:2002ci}
  P.~de Forcrand and O.~Philipsen,
  Nucl.\ Phys.\ B {\bf 642} (2002) 290
  [hep-lat/0205016];
  Nucl.\ Phys.\ B {\bf 673} (2003) 170
  [hep-lat/0307020].

\bibitem{Toublan:2005rq}
  D.~Toublan,
  Phys.\ Lett.\ B {\bf 621} (2005) 145
  [hep-th/0501069].

\bibitem{Braun:2011iz}
  J.~Braun, B.~Klein and B.~-J.~Schaefer,
  arXiv:1110.0849 [hep-ph].

\bibitem{Schaefer:2004en}
  B.~-J.~Schaefer and J.~Wambach,
  Nucl.\ Phys.\ A {\bf 757} (2005) 479
  [nucl-th/0403039].

\bibitem{Strodthoff:2011tz}
  N.~Strodthoff, B.~-J.~Schaefer and L.~von Smekal,
  arXiv:1112.5401 [hep-ph].
 
\bibitem{Kojo:2011cn}
  T.~Kojo, Y.~Hidaka, K.~Fukushima, L.~McLerran and R.~D.~Pisarski,
  Nucl.\ Phys.\ A {\bf 875} (2012) 94
  [arXiv:1107.2124 [hep-ph]].

\bibitem{Carignano:2010ac}
  S.~Carignano, D.~Nickel and M.~Buballa,
  Phys.\ Rev.\ D {\bf 82} (2010) 054009
  [arXiv:1007.1397 [hep-ph]].


  
\end{thebibliography}
\end{document}